\documentclass[entropy,article,submit,moreauthors,pdftex,10pt,a4paper]{mdpi} 
\firstpage{1} 
\articlenumber{x}
\doinum{10.3390/------}
\pubvolume{xx}
\pubyear{2017}
\copyrightyear{2017}
\externaleditor{Academic Editor: name}
\history{Received: date; Accepted: date; Published: date}

\Title{Stochastic Stirling engine operating in contact with active baths}

\Author{Ruben Zakine $^{1}$, Alexandre Solon $^{2}$, Todd Gingrich $^{2}$ and Fr\'ed\'eric van Wijland $^{1}$}

\AuthorNames{Ruben Zakine, Alexandre Solon, Todd Gingrich and Fr\'ed\'eric van Wijland}

\address{%
$^{1}$ \quad Universit\'e Paris Diderot, Sorbonne Paris Cit\'e, MSC, UMR 7057 CNRS, 75205 Paris, France\\
$^{2}$ \quad Department of Physics, Massachusetts Institute of Technology, Cambridge, MA 02139, USA}

\corres{Correspondence: fvw@univ-paris-diderot.fr}

\abstract{A Stirling engine made of a colloidal particle in contact with a nonequilibrium bath
is considered and analyzed with the tools of stochastic energetics. We model the bath by non
Gaussian persistent noise acting on the colloidal particle. Depending on the chosen definition of
an isothermal transformation in this nonequilibrium setting, we find that either the energetics of the
engine parallels that of its equilibrium counterpart or, in the simplest case, that it ends up being less
efficient. Persistence, more than non Gaussian effects, are responsible for this result.
}


\newcommand{\dd}{\text{d}}

\newcommand{\ee}{\text{e}}

\newcommand{\p}{\partial}

\newcommand{\br}{\text{\bf r}}

\renewcommand{\vec}[1]{\mathbf{#1}}

\newcommand{\grad}{\nabla}
\newcommand{\cP}{\mathcal{P}}

\begin{document}

\section{Introduction}

Every well-educated physicist has heard of Carnot or Stirling
cycles. In equilibrium thermodynamics of macroscopic systems (such as
a gas enclosed in some container), a cycle is a periodic sequence of
transformations the system is subjected to, with a view, as far as
engines are considered, to extracting work from the system. For a
Carnot cycle, this is the well-known
adiabatic-isothermal-adiabatic-isothermal sequence, while for a
Stirling cycle the adiabatic transformations are replaced with isochoric
ones. The analysis of small, microscopic or nanoscopic
systems, such as a colloidal particle in some solvent, by contrast to
the nineteenth century fluid systems, poses theoretical and
experimental challenges. The former have been overcome by the advent
of stochastic energetics at the end of the
nineties~\cite{Sekimoto-book}. Stochastic energetics (or stochastic
thermodynamics) encompasses a series of concepts and methods that
allow one to define work, heat, dissipation, energy, {\it etc.} at an
instantaneous and fluctuating level. By taking averages one usually
recovers (with often no need to consider the limit of macroscopic
systems) the standard principles of thermodynamics. The gain, however,
is enormous in that stochastic energetics also allows one to quantify
fluctuations, which may not be negligible for small-scale systems. An
excellent review on the latest developments of stochastic
thermodynamics is that by Seifert~\cite{0034-4885-75-12-126001}, while
the earlier Schmiedl and Seifert paper~\cite{0295-5075-81-2-20003}
focuses specifically on the analysis of stochastic
engines. Experimental realizations pose challenges of their own. These
are concerned with the control of small-size objects (often by means
of optical tweezers), coupled to the need to control other parameters
of the experiment. The bath temperature is one of them. Another one is
the optical trap stiffness that can be seen as playing a role
analogous to the volume of the container enclosing the gas in the
macroscopic version. The conjugate parameter (analogous to the
pressure) is the particle position (squared). The first colloidal-made
engines were concerned with a Stirling
cycle~\cite{blickle_realization_2011,horowitz_thermodynamics:_2012},
in which a sequence of transformations by which the bath temperature
and the trap stiffness were varied was applied to the colloidal
particle. This is no place to discuss what an adiabatic transformation
actually means at the level of a colloidal particle in a solvent,
suffice it to say that this has very recently been
defined~\cite{PhysRevLett.114.120601} and put to work in an actual
Carnot cycle~\cite{martinez_brownian_2016}. A lot remains to be done
at the experimental level and theoretical level alike, but it is fair
to say that things are pretty well-understood as far as the
theoretical framework is concerned. However, a somewhat unexpected
generalization of these cycles seen as transformations between
equilibrium states has recently been put forward by Krishnamurthy {\it
et al.}~\cite{krishnamurthy_micrometre-sized_2016}. The
generalization, in the spirit of the seminal work of Wu and
Libchaber~\cite{PhysRevLett.84.3017}, consists in replacing the
equilibrium bath by an active bath containing living bacteria in a
stationary yet nonequilibrium state. The sequence of transformations
thus occurs between nonequilibrium steady-states instead of between
equilibrium ones. Due to the nonequilibrium nature of the bacterial
bath, there is no way to define a {\it bona fide} temperature. There
are, however, several ways to define an energy scale expressing the
level of energetic activity of the bath (all of which reduce, in some
equilibrium limit, to the physical temperature). The proposal of
\cite{krishnamurthy_micrometre-sized_2016} is to use the colloid's
position fluctuations via $T_\text{act}=\frac{k}{2}\langle x^2\rangle$
(where $k$ is the trap stiffness). Another posssibility would have
been the following: in the absence of any confining force, the
colloidal particle will eventually diffuse away from its initial
position, so that we might then expect $\langle
(x(t)-x(0))^2\rangle=\frac{2T}{\gamma} t$, where $T$ is yet another
acceptable active temperature (this would be the asymptotic slope in
figure 2 of \cite{PhysRevLett.84.3017}). One might be inclined,
somewhat subjectively, to view $T$ as better expressing the intrinsic
properties of the bath, while $T_\text{act}$ must result from a
balance between the bath and some external force. We will come back to
that point at a later stage.

The purpose of this work is to analyze the results of
\cite{krishnamurthy_micrometre-sized_2016} in the light of a specific
modeling of the bacterial bath. Our modeling relies on a single
hypothesis: the bath enters the colloid's motion only through an extra
noise term, and the noise statistics alone encode for the effect of
the bath. Inspired by the suggestion of
\cite{krishnamurthy_micrometre-sized_2016} that non Gaussian
statistics are essential, we will propose that the noise to which the
colloid is subjected may have itself non Gaussian statistics (recent
advances of stochastic energetics for non Gaussian but white
processes~\cite{PhysRevLett.108.210601,kanazawa2013heat} have taught
us how to manipulate such signals) and possibly possess persistence
properties. We will begin by a reminder of the properties of the
stochastic Stirling engine between equilibrium reservoirs. We will
then consider the extension to nonequilibrium bath and see how
equilibrium results are {\it not} affected by choosing isothermal
processes based on $T_\text{act}$. Then, we will adopt a definition of
active temperature based on the colloid's diffusion constant and show
that energy balance considerations are deeply modified and that the
persitence of the noise is of key importance.

\section{Stirling cycle between between equilibrium states: a quick review}
\subsection{Modeling the motion of a colloidal particle}
The standard description of the dynamics of a colloidal particle in a
solvent rests on a Langevin equation governing the evolution of the
particle's position $x(t)$. In the overdamped limit relevant to the
description of a micron-sized particle, this Langevin equation reads
\begin{equation}\label{eq-Langevin}
\gamma\frac{\dd x}{\dd t}=-\p_x V+\gamma\eta
\end{equation}
where $\gamma$ is the friction coefficient characterizing the viscous
drag of the particle in the solvent (this is the inverse
mobility). The external potential $V$ depends on the particle's
position $x$ and an external control parameter of the potential (like
the stiffness of the harmonic trap). Finally, $\eta$, which, with the
chosen normalization, has the dimension of a velocity, stands for a
Gaussian white noise with correlations
$\langle\eta(t)\eta(t')\rangle=\frac{2T}{\gamma}\delta(t-t')$.  Under
those conditions, where the dissipation kernel exactly matches the
noise correlator, as prescribed by Kubo~\cite{0034-4885-29-1-306}, the
colloidal particle is in equilibrium (provided, of couse, the external
potential is not time dependent). In experimental setups, the
potential is harmonic and the particle's motion is tracked in
two-dimensional space, $\br=(x,y)$ and $V(x,k)=\frac{k}{2} (x^2+y^2)$.
We will stick to a one-dimensional description for notational
simplicity. In the nonequilibrium setting we want to describe here, we
shall encapsulate the effects of the interactions of the colloidal
particle with its nonequilibrium environment into a single ingredient,
namely the noise statistics. But there is no reason to expect the
noise resulting from the interactions of the colloidal particle with
the bacteria bath to be either Gaussian or white. We postpone the
analysis of such active noises to the next section and now proceed
with a reminder of
\cite{0295-5075-81-2-20003,blickle_realization_2011}.

\subsection{Energetics of the Stirling cycle}\label{Stirling-eq}
\begin{figure}[H]
\centering
\includegraphics[width=10cm]{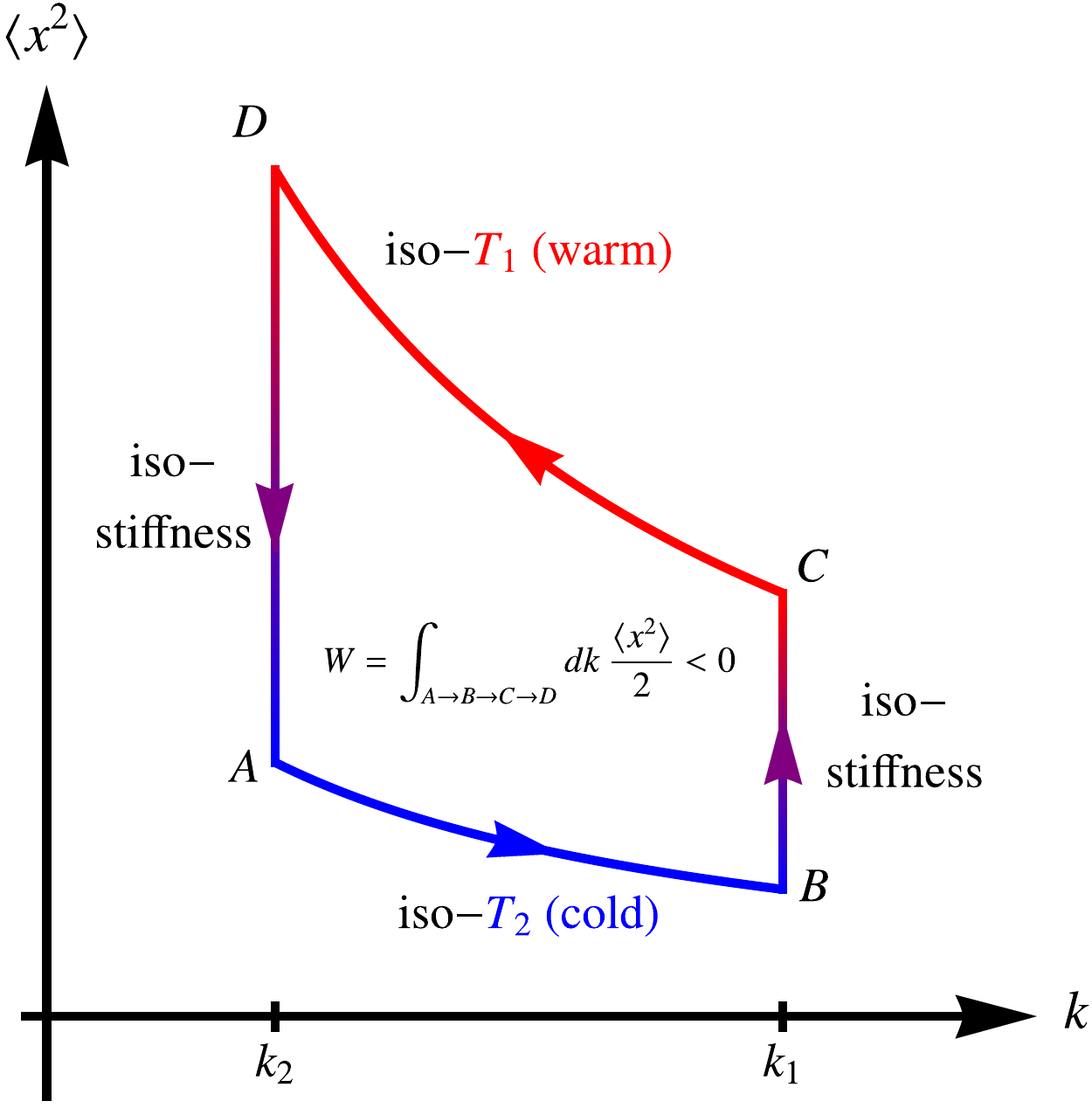}
\caption{\label{sketch}Schematic diagram of the Stirling cycle in stiffness-position space. Unlike its thermodynamic counterpart, the cycle is run counter-clockwise but is nevertheless an engine cycle.}
\end{figure}

In this subsection, we briefly review the results presented in
\cite{blickle_realization_2011,horowitz_thermodynamics:_2012}. This
serves as a way to set notations straight and to define the quantities
of interest. A Stirling cycle $ABCDA$ is made of the following
sequence of states in the stiffness-temperature space $(k,T)$:
\begin{equation}\label{cycle}
A:\;(k_2,T_2)\overset{\text{isothermal}}{\longrightarrow} B:\; (k_1,T_2)\overset{\text{isochoric}}{\longrightarrow} C:\; (k_1,T_1)\overset{\text{isothermal}}{\longrightarrow} D:\; (k_2,T_1)\overset{\text{isochoric}}{\longrightarrow} A
\end{equation}
where the terminology {\it isochoric} of course refers to an
iso-stiffness transformation. This cycle is sketched in figure
\ref{sketch}.
We will denote by $a=k_1/k_2>1$ the stiffness ratio (a large value of
$k$ is analogous to a more compressed state). The warm source is at
$T_1$ while the cold source is at $T_2$ ($T_1>T_2)$. The instantaneous
fluctuating energy of the particle is $V(x,k)=\frac{k}{2}x^2$. The
work done on the colloid along a protocol driving it from state $i$
to state $f$ is
$W=\int_i^f\dd t\frac{\dd k}{\dd t}\frac{\p V}{\p k}=\int_i^f\dd
k\frac{1}{2} x^2$.
The heat received by the colloid during the same step is given by the
integral of the entropy production along the given protocol:
\begin{equation}\label{heat-entropy-prod} Q=-\int_i^f\dd t T\sigma
\end{equation} where
$\sigma=T^{-1}\dot{x}(\gamma\dot{x}-\gamma\eta)=-T^{-1}k\dot{x}x$ is
also the rate of work performed by the particle on the bath, and thus $Q$ is
the work performed by the bath on the particle. Altogether we thus
have $Q=\int_i^f k x\dd x$. If $p_\text{eq}(x)=\ee^{- k
x^2/2T}/\sqrt{2\pi T/k}$ is the equilibrium distribution then, up to a
constant $S=-\int \dd x p_\text{eq}(x)\ln p_\text{eq}(x)=-\frac 12
\ln\frac{k}{2\pi T}+\frac 12$ is the equilibrium entropy and $\langle
Q\rangle =\int_i^f T\dd S$, with $\dd S=-\frac 12 \frac{\dd
k}{k}+\frac 12 \frac{\dd T}{T}$. This is consistent with $Q=\int_i^f k
x\dd x=[\frac{k x^2}{2}]_i^f-\int_i^f \dd k \frac{x^2}{2}$, which is a
promotion of the first law $V_f-V_i=W+Q$ to stochastic
energies. Using that $\langle x^2\rangle =T/k$, it is a simple
exercise to determine the average heat received by the system during each
step, $\langle Q_{AB}\rangle=-\frac 12 T_2\ln a<0$, 
$\langle Q_{BC} \rangle=\frac 12 (T_1-T_2)>0$,
$\langle Q_{CD} \rangle=\frac 12 T_1 \ln a>0$ and
$\langle Q_{DA} \rangle =-\frac{1}{2}(T_1-T_2)<0$. Correspondingly,
$\langle W_{AB} \rangle=\frac{T_2}{2}\ln a$, $\langle W_{BC}\rangle =0$, 
$\langle W_{CD}\rangle =-\frac{T_1}{2}\ln a$
and $\langle W_{DA}\rangle=0$. The total average work received by the 
colloid is  $\langle W \rangle = \langle W_{AB}+W_{CD} \rangle=-\frac{1}{2}(T_1-T_2)\ln a<0$. This means that the
engine provides some work on average. Given that $Q_1=Q_{BC}+Q_{CD}$ and
$Q_2=Q_{AB}+Q_{DA}$ are the heat effectively received by the colloid
and the heat effectively given by the colloid to the bath,
respectively, we define ${\mathcal E}=\frac{|\langle W\rangle |}{\langle Q_1\rangle}$ as the engine's
efficiency. The result
\begin{equation}\label{efficiency-eq} {\mathcal
E}=\frac{\langle Q_1+Q_2 \rangle}{\langle Q_1 \rangle}=\frac{(T_1-T_2)\ln a}{T_1-T_2+T_1\ln a}
\end{equation} If a perfect regenerator was used during the isochoric
cooling $D\to A$ then the energy given out during this isochoric
cooling could be used for the heating during the isochoric heating
$B\to C$. Then the heat received by the colloid would reduce to
$Q_1=Q_{CD}$ and the efficiency would become ${\mathcal
  E}_C=\frac{(T_1-T_2)\ln a}{T_1\ln a}=1-\frac{T_2}{T_1}$ (this Carnot
efficiency is of course an upper bound for ${\mathcal
  E}=\frac{{\mathcal E}_C\ln a}{{\mathcal E}_C+\ln a}$ as given in
\eqref{efficiency-eq}). Again, these results can all be found in
\cite{blickle_realization_2011}. We have now set the stage for the
purpose of this work, which is to re-examine each of these steps when
the colloidal particle is in contact with nonequilibrium baths just as
was carried out experimentally in \cite{krishnamurthy_micrometre-sized_2016}.

\section{Engine operating between nonequilibrium baths}
\subsection{Modified Langevin equation} We stick to our hypothesis
\label{sec:modifiedLE}
that the effects of the bacterial bath can be entirely encoded into a
single random process, so that now the colloid's position evolves
according to
\begin{equation} \gamma\dot{x}=-k x+\gamma\eta_\text{act}
\end{equation} where the active noise $\eta_\text{act}$ is a
characteristic feature of the bacterial bath. Assuming this random
signal inherits its properties from the bacteria making up the bath,
we may expect that not only will the noise display non Gaussian
statistics but it will also exhibit persistence properties captured by
some memory kernel in the noise correlations. Among existing models,
we may cite Run-and-Tumble noise, Active Brownian noise (see
\cite{solonEPJST} for a review), Active Ornstein-Uhlenbeck
noise~\cite{szamel2014self} or even white yet non Gaussian
\cite{PhysRevLett.108.210601}. Following
\cite{krishnamurthy_micrometre-sized_2016} we define a first active
temperature $T_\text{act}$ by the steady-state value of $x^2$:
$T_\text{act}\equiv k\langle x^2\rangle$. However, we introduce
another active temperature that we denote by $T$ by means of the
colloid's mean-square displacement in the absence of a confining
force, namely at $k=0$ we expect that
\begin{equation}\label{def-T} \langle (x(t)-x(0))^2\rangle=\frac{2
T}{\gamma} t
\end{equation} at large times, so that $T=\frac{\gamma}{2 t}\int_0^t
\dd t_1\dd t_2 \langle
\eta_\text{act}(t_1)\eta_\text{act}(t_2)\rangle$. We stress that
neither $T$ nor $T_\text{act}$ are {\it bona fide} temperatures. They
merely are energy scales reflecting how the bath injects energy into
the colloids.
\subsection{The energetics is not altered if we use iso-$T_\text{act}$
  steps} Using the definition of the work
$W_{i\to f}=\int_i^f\dd k\frac{x^2}{2}$ we see that
$\langle W_{i\to f}\rangle=\int_i^f\dd k\frac{T_\text{act}}{2 k}$
which leads to the exact same expressions for the work as found in the
previous section, up to the replacement of the equilibrium temperature
with $T_\text {act}$. Similarly, the heat is given by the work exerted
by the bath on the colloid, namely
$Q_{i\to f}=\int_i^f\dd
t\dot{x}(-\gamma\dot{x}+\gamma\eta_\text{act})$,
which again simplifies into $Q_{i\to f}=\int_i^f\dd t\dot{x}(k x)$ and
thus $Q_{i\to f}=[\frac{k x^2}{2}]_i^f-\int_i^f\dd k\frac{x^2}{2}$.
After taking averages, we are back onto the expression found in
equilibrium, again up to the replacement of temperatures by the
corresponding $T_\text{act}$'s. Hence, within that set of definitions
and within our modeling, a quasistatic engine operating between
nonequilibrium baths cannot outperform an equilibrium one.  In light
of the experiments of \cite{krishnamurthy_micrometre-sized_2016} this
leaves us with a puzzle that we will adress in the discussion
section. In the following section, we suggest that perhaps another
definition of the active isothermal process might lead to more
striking differences with respect to an equilibrium engine.

\section{Energetics using the diffusion constant as an active temperature}
In this section we re-examine the Stirling engine operating between
nonequilibrium baths using the temperature $T$ defined in
Eq.~\eqref{def-T} via the diffusion constant of an unconstrained
particle. An isothermal process will now be understood as a process at
constant $T$. Physically, this requirement is arguably more natural
than processes at constant $T_{\rm act}$. Indeed, $T$ is an intrinsic
measure of the activity of the bath, which can usually be tuned easily
by the experimentalist, while $T_{\rm act}$ results from a balance
between the bath's activity and a given external potential.

This new definition immediately requires us to adopt specific models
for the active noise $\eta_\text{act}$ because the explicit dependence
of $\langle x^2\rangle$ on $T$ and $k$ is now of crucial
importance. We examine successively the case in which
$\eta_\text{act}$ is a non Gaussian but white noise, then an
persistent noise with two-point correlations exponentially decreasing
in time, a case that encompasses Ornstein-Uhlenbeck noise,
Run-and-Tumble or Active Brownian noise.

\subsection{A bath with white but non Gaussian statistics}
\label{Sec:nonGaussian}
Let's now assume that the active nature of the bacterial bath only
surfaces through the non Gaussian statistics of the noise
$\eta_\text{nG}$ appearing in the Langevin equation,
\begin{equation}
\label{Langevin-ng} \gamma\dot{x}=-\p_x V+\gamma\eta_\text{nG}
\end{equation} 
while memory effects can be ignored in a first
approximation. A non Gaussian white noise $\eta_\text{nG}(t)$ can be
formed by compounding Poisson point processes with random and independent 
amplitudes\cite{van1992stochastic}. In practice, a realization of the noise
over a time interval $[0,\mathcal{T}]$ is generated by first drawing
a number of points, $n$, from a Poisson distribution with mean 
$\nu \mathcal{T}$. Then a collection of times $t_i$, with $i=1,\ldots,n$, 
are drawn uniformly in $[0, \mathcal{T}]$.
To each $t_i$ is associated a jump amplitude $c_i$, where the $c_i$'s are 
independent but identically distributed random variables with distribution 
$p(c)$. 
The non Gaussian, white noise is constructed as the composition of these
random-amplitude Poisson jumps:
\begin{equation}
\eta_\text{nG}(t)=\sum_ic_i\delta(t-t_i)
\end{equation}
The generating functional of $\eta_\text{nG}(t)$ is
$\langle\ee^{\int \dd t \;j(t)\eta_\text{nG}(t)}\rangle=\ee^{\nu\int\dd
  t(\langle\ee^{c \;j(t)}\rangle_p-1)}$,
where the $p$ index denotes an average with respect to $c$ and $j(t)$ is
the field conjugate to $\eta_{\text{nG}}$. The two
parameters defining the noise statistics are the hitting frequency
$\nu$ and the full jump distribution $p$. The Gaussian white noise
limit is recovered as $\nu\to\infty$ and $\langle c^2\rangle_p\to 0$
while $\nu\langle c^2\rangle_p$ remains fixed. The noise has cumulants
\begin{equation} 
\langle \eta_\text{nG}(t_1)\ldots\eta_\text{nG}(t_n)\rangle_\text{cumulant}=\nu\langle
c^n\rangle_p\delta(t_1-t_2)\ldots \delta(t_{n-1}-t_n)
\end{equation}
We denote by $T/\gamma=\nu \langle c^2\rangle_p/2$ so
that $\langle \eta_\text{nG}(t)\eta_\text{nG}(t')
\rangle=(2T/\gamma)\delta(t-t')$ and $T$ matches the definition given
in Eq.~\eqref{def-T}. 
It is possible to show that, in the case of the non Gaussian white noise, 
this $T$ is actually identical to our prior definition 
$T_\text{act} = \langle x^2\rangle/k$.
To prove this, we start from the master equation for the probability that $x(t)$ takes the value $x$ at time $t$, $P(x,t)$, which reads
\begin{equation}
\p_tP(x,t)=\gamma^{-1}\p_x(k xP(x,t))+\nu\int\dd c\;p(c)\left(P(x-c,t)-P(x,t)\right)
\end{equation}
which we multiply by $x^2$ and integrate over $x$. This directly leads to 
\begin{equation}\begin{split} \frac{\dd}{\dd t}\langle
x^2\rangle=&-\frac{2k}{\gamma}\langle x^2\rangle+ \nu \langle c^2\rangle_p
\end{split}\end{equation}
Hence it results that in steady-state
$k\langle x^2\rangle=T$, independently of the non Gaussian noise
specifics. This is an extension of equipartition to
a nonequilibrium context~\footnote{An identical equipartition holds
for an underdamped Langevin equation with non Gaussian noise, for
which $m\dot{v}=-\gamma v-V'+\gamma\eta_\text{nG}$ leads to
$m\frac{\dd\langle v^2\rangle}{\dd t}=-\gamma\langle
v^2\rangle-\frac{\dd}{\dd t}\langle V\rangle+\frac{2\gamma
T}{m}$. This indeed forces $\langle mv^2/2\rangle=T/2$ in the
steady-state, irrespective of the white noise statistics. In a similar
vein, one can also see that $\frac{\dd}{\dd t}\langle
xv+\frac{\gamma}{m} \frac{x^2}{2}\rangle=\langle
v^2\rangle-\frac{1}{m}\langle x V'\rangle$, which leads to $\langle x
V'\rangle=m\langle v^2\rangle=T$ in the nonequilibrium
steady-state.}. It immediately follows that the average works and
heats will be unchanged with respect to the equilibrium discussion of
subsection \ref{Stirling-eq}. Note, however, that the heat is not
given anymore by the entropy production $\delta Q=-T\sigma\dd t$ as in
Eq.~\eqref{heat-entropy-prod}. It would be an interesting task to try
and evaluate the corresponding $\sigma$ which is beyond the scope of
the present discussion. (This might be feasible in an expansion in the
jump size $a$ at fixed active $T$. Such an expansion around a Gaussian
white noise is admittedly questionable in view of Pawula's
theorem~\cite{1053955} but can be controled when manipulated with
care~\cite{popescu2015kramers}). Finally, that equipartition holds
does not preclude strong non Gaussian effects to show up in the
colloid's position pdf. For instance, choosing
$p(c)=\ee^{-|c|/a}/(2a)$ (with $\langle c^2\rangle_p=2 a^2$) for the
distribution of jumps allows us to find the stationary state
distribution $P_{\text{ss}}(x)$. Indeed, with this choice of jump
statistics~\cite{fodor-kyoto} for $\alpha=\frac{T}{2 ka^2}-\frac 12$
positive we have $P_{\text{ss}}(x)=C|x/a|^{\alpha}K_\alpha(|x|/a)$
and $C=2^{-\alpha} a^{-1}/(\sqrt{\pi}\Gamma(1/2+\alpha))$, and thus
$\langle x^2\rangle=\frac{T}{k}$ can be explicitly verified. Note
$P_\text{ss}$ exhibits a cusp at the origin for $0\leq\alpha\leq
1/2$, namely for $ka^2<T<2ka^2$.
\begin{figure}[H] \centering
\includegraphics[width=10cm]{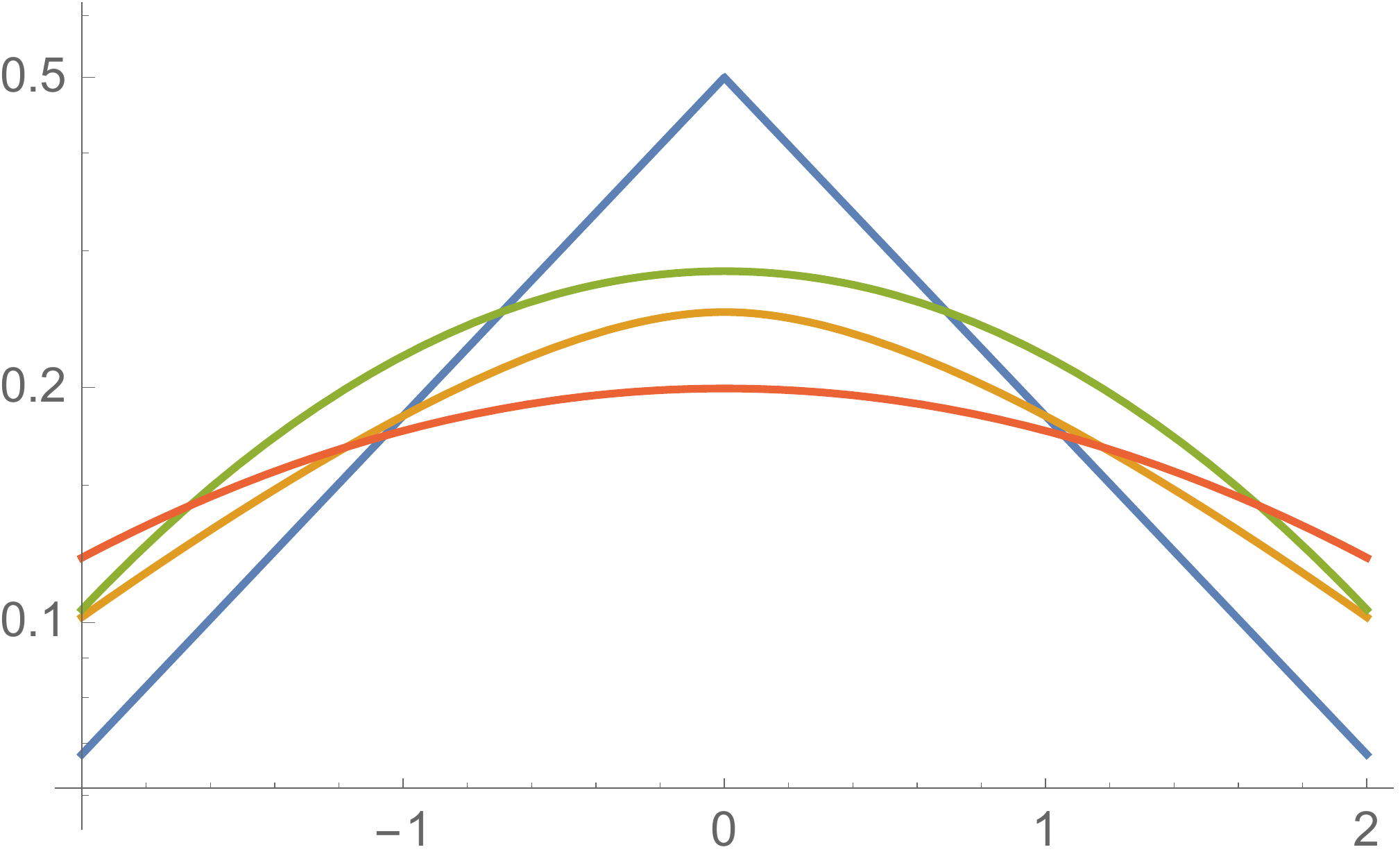}
\caption{\label{fig-pdf}Log of the probability of the colloid's position as a
function of position (for a unit $a$), in equilibrium with Gaussian
statistics (red at $T/k=2$, green at $T/k=4$) or out of equilibrium as
given by $P_\text{ss}$ (blue at $T/k=2$, orange at $T/k=4$).}
\end{figure} It comes as no surprise that the position statistics in
the steady-state are strongly non Gaussian as illustrated in figure
\ref{fig-pdf}. However, it simply turns out that these non Gaussian
fluctuations do not interfere with the energy balance of the Stirling
engine (Appendix~\ref{appendix:kurtosis} shows explicitly that the
value of the kurtosis of the position distribution is uncorrelated
from the efficiency).  We now turn our attention to an active noise
displaying some persistence properties with however Gaussian
statistics.

\subsection{A bath with a persistent noise}
\label{bath-pers} We now address more realistic modelings of the noise
produced by the bacterial bath, in the form of a stochastic
force imparted on the colloid that captures the persistent motion of
an active particle. Such persistent noise arises from three common classes
of active dynamics: Run-and-Tumble particles, active Brownian particles,
and active Ornstein-Uhlenbeck motion. All three classes exhibit noise
correlations that decay exponentially with a characteristic time $\tau$:
\begin{equation}\label{noiseP}
\langle\eta_\text{P}(t)\eta_\text{P}(t')\rangle=\frac{2T}{\gamma}\times\frac{\ee^{-\frac{|t-t'|}{\tau}}}{2\tau}.
\end{equation}
The prefactor $T$ in Eq.~\eqref{noiseP} matches our definition for $T$ from 
Eq.~\eqref{def-T}. We show in the Appendix~\ref{sec:appActivedyn} how the different models
give rise to Eq.~(\eqref{noiseP}) and relate $T$ and $\tau$ to the
microscopic parameters of the dynamics. Here we adopt a unified description
of the three different models by analyzing the impact of their shared noise
correlator, Eq.~\eqref{noiseP}. Note that we restrict our discussion to one
space dimension only for simplicity. In higher dimensions, the correlator
of each component of the (vectorial) noise is still given by 
Eq.~\eqref{noiseP}, and, by symmetry, our results trivially generalize to a
spherically harmonic potential.

If we interpret the isothermal transformations of Fig.~\ref{sketch} as iso-$T$
processes (as opposed to iso-$T_\text{act}$), the energetics of our
Sirling cycle now differs from the equilibrium analysis of subsection
\ref{Stirling-eq}. During an iso-$T$ protocol, $\langle x^2\rangle$ does not 
trace an isotherm with the form $\langle x^2\rangle \propto k^{-1}$. 
The new form of the isotherm depends only on the two-point correlator
$\langle\eta_\text{P}(t)\eta_\text{P}(t')\rangle$ and not on
higher-order correlations, allowing us to simultaneously treat all three
types of active motion. Indeed, in Fourier space, Eq.~\eqref{Langevin-ng} reads
\begin{equation}
  \label{eq:dynamics-Fourier}
  (k+i\gamma\omega)\tilde x(\omega)=\gamma\tilde\eta_{\rm P}(\omega)
\end{equation}
with the Fourier transform defined as
$\tilde f(\omega)=\int_{-\infty}^{+\infty}f(t)e^{-i\omega t}\dd t$. One
can then show that in steady state
\begin{equation}
  \label{eq:x2-correlator}
  \langle x^2\rangle =\int_{-\infty}^{+\infty}\frac{\dd \omega}{2\pi} \frac{\gamma^2\langle \tilde\eta_{\rm P}(\omega)\tilde\eta_{\rm P}(-\omega)\rangle}{k^2+\omega^2\gamma^2}.
\end{equation}
For the noise correlator Eq.~\eqref{noiseP},
$\langle \tilde\eta_{\rm P}(\omega)\tilde\eta_{\rm
  P}(-\omega)\rangle=2T/(\gamma(1+\tau^2\omega^2))$
so that we obtain
\begin{equation}
  \label{eq:x2-explicit}
  \langle x^2\rangle =\frac{T}{k(1+k\tau/\gamma)}.
\end{equation}
In the notation of Sec.~\ref{sec:modifiedLE},
$T_{\rm act}\equiv k \langle x^2 \rangle =T/(1+\Omega \tau)$ where we
have defined the frequency $\Omega\equiv k/\gamma$.

Let us proceed, then, with the cycle Eq.~\eqref{cycle} in which we
consider isothermal processes at fixed $T$. The average work has the
expression
$\langle W_{i\to f}\rangle=\frac{1}{2}\int_i^f\dd k\frac{T}{k(1+\Omega\tau)}$,
which is zero along an isochoric protocol, but which now reads
$\langle W_{i\to f}\rangle=\frac{T}{2}[\ln
\frac{k}{1+\Omega\tau}]_i^f$
along an iso-$T$ protocol. Similarly, the average heat reads
$\langle Q_{i\to f}\rangle=[\frac{T/2}{1+\Omega\tau}]_i^f-
\frac{1}{2}\int_i^f\dd k\frac{T}{k(1+\Omega\tau)}$. 
Putting everything together we find
\begin{equation}
\begin{aligned}
\langle Q_{AB}\rangle&=\frac{T_2}{2}\left[\frac{1}{1+\Omega_1\tau}-\frac{1}{1+\Omega_2\tau}\right]-\frac{T_2}{2}\ln\left[a\frac{1+\Omega_2\tau}{1+\Omega_1\tau}\right]
\\ 
\langle Q_{BC}\rangle&=\frac{(T_1-T_2)/2}{1+\Omega_1\tau}>0 \\
\langle Q_{CD}\rangle&=\frac{T_1}{2}\left[\frac{1}{1+\Omega_2\tau}-\frac{1}{1+\Omega_1\tau}\right]+\frac{T_1}{2}\ln\left[a\frac{1+\Omega_2\tau}{1+\Omega_1\tau}\right]
\\ 
\langle Q_{DA}\rangle&=-\frac{(T_1-T_2)/2}{1+\Omega_2\tau}<0
\end{aligned}\end{equation} 
while the average works are given by
\begin{equation} \langle W_{AB}\rangle=\frac{T_2}{2}\ln
a\frac{1+\Omega_2\tau}{1+\Omega_1\tau},\; 
\langle W_{CD}\rangle=-\frac{T_1}{2}\ln a\frac{1+\Omega_2\tau}{1+\Omega_1\tau}
,\; \langle W_{BC}\rangle = \langle W_{DA}\rangle =0
\end{equation}
and thus ${\mathcal E}=\frac{-\langle W\rangle}{\langle Q_{BC}+Q_{CD}\rangle}$ in the $T_1\gg T_2$ limit is
\begin{equation} {\mathcal E}_\text{sat}=\frac{\ln
a\frac{1+\Omega_2\tau}{1+\Omega_1\tau}}{\ln
a\frac{1+\Omega_2\tau}{1+\Omega_1\tau}+\frac{1}{1+\Omega_2\tau}}
\end{equation} 
In the limit of small correlation time ($\Omega_1 \tau \ll 1$), we find
that 
\begin{equation}\label{eff-sat} {\mathcal E}_\text{sat}\simeq\frac{\ln
a}{1+\ln a}-\Omega_2\tau\frac{a-1-\ln a}{(1+\ln a)^2}+{\mathcal
O}(\tau^2).
\end{equation} 
In Eq.~\eqref{eff-sat} the correction to $\mathcal{E}_\text{sat}$ actually 
remains negative at arbitray values of $\tau$:
the efficiency saturates to a {\it lower} value due to the persistent
properties of the noise when compared to equilibrium (no
persistence). The cost of maintaining nonequilibrium steady-state is
not paid off by an improved efficiency! While the available work has
increased, the required energy to operate the engine has increased by
an even larger amount.\\

Let us stress here that the generality of these results, which depend
only on the two-point correlator of the noise but not on higher-order
statistics, can seem rather surprising because the behavior of active
Ornstein-Uhlenbeck, Run-and-Tumble and Active Brownian particles in an
harmonic trap are all qualitatively different. The case of an active
Ornstein-Uhlenbeck noise in a quadratic potential is special in that
the colloidal particle actually {\it is} in
equilibrium~\cite{szamel2014self}. The equilibrium distribution is
$p_\text{eq}(x)\sim\ee^{-\frac{k(1+\Omega\tau)}{2T}x^2}$ (whether and
how this extends beyond a quadratic potential was discussed in
\cite{fodor2016far}), from which
$T_{\rm act}=T/(1+\Omega \tau)$ is readily extracted. On the
contrary, Active Brownian and Run-and-Tumble particles have a richer
(nonequilibrium) physics. In particular, if the particle is persistent
on a time scale larger than $\Omega^{-1}$, the steady-state distribution
is not peaked around $x=0$ (the particle spends most of its time on
the edge of the trap)~\cite{solonEPJST}. It is therefore surprising
that these differences do not affect the thermodynamics of our engine.

\subsection{A bath described by a more general Langevin equation}
A more general description of the effect of the bath on the colloidal
particle includes memory effect in the dissipation as well, as
discussed by Berthier and Kurchan~\cite{berthier2013non} in a
different context. One obtains a Langevin equation of the form
\begin{equation}
  \label{eq:Langevin-general} \gamma \int_{-\infty}^t\dd
t'K_\text{diss}(t-t')\dot{x}(t')=-kx +\gamma\eta;\quad
\langle\eta(t)\eta(t')\rangle=\frac{2T}{\gamma}K_\text{inj}(|t-t'|)
\end{equation} with generic injection and dissipation kernels
$K_\text{inj}$ and $K_\text{diss}$. Equilibrium is achieved on
condition that $K_\text{inj}(\omega)$ and $\text{Re}
K_\text{diss}(\omega)$ are equal. The ratio
\begin{equation}
T_\text{eff}(\omega)=T\frac{K_\text{inj}(\omega)}{\text{Re}
K_\text{diss}(\omega)}
\end{equation} tells us about the mismatch between injection and
dissipation. In the cases considered previously, with
$K_\text{diss}(\omega)=1$, we ended up with $T_{\text{eff}}(\omega)=\frac{T}{1+(\omega\tau)^2}$. The characteristic
frequency of the relaxation within the harmonic well being $\Omega$,
we {\it a posteriori} understand that in the regime where persistence
matters, namely when $\Omega\tau\gg 1$,
$T_\text{eff}(\omega>\Omega)\to 0$, and thus the position spectrum
will be cut-off beyond $\omega=\Omega$. It is thus no surprise that
eventually $\langle x^2\rangle<T/k$, or $T_\text{act}<T$. Of course,
in the presence of a more complicated dissipation kernel
$K_\text{diss}$, the latter inequality can be challenged. Should one
devise a dissipation kernal such that $T_\text{act} > T$, one might
reasonably suspect that the thermodynamic efficiency could exceed that
of the equilibrium Stirling engine.\\

Let us therefore consider how the energetics of the cycle in
Fig.~\ref{sketch} is impacted by the relationship between
$T_\text{act}$ and $T$.  In particular, we repeat the analysis of
Sec.~\ref{bath-pers} by assuming $T_\text{act} = T f(k)$ for a generic
function $f$. The derivation proceeds exactly as before and we obtain
for the heat and work on each segment
\begin{align}
  \langle W_{AB}\rangle=\frac{T_2}{2}\left[G(k_1)-G(k_2)\right];\quad \langle W_{CD}\rangle=-\frac{T_1}{2}\left[G(k_1)-G(k_2)\right];\quad \langle W_{BC}\rangle=\langle W_{DA}\rangle=0 \\
  \langle Q_{AB}\rangle=-\frac{T_2}{2}\left[G(k_1)-G(k_2)\right]+\frac{T_2}{2}\left[f(k_1)-f(k_2)\right];\quad\langle Q_{BC}\rangle=\frac{(T_1-T_2)}{2}f(k_1); \\
 \langle Q_{CD}\rangle=\frac{T_1}{2}\left[G(k_1)-G(k_2)\right]-\frac{T_1}{2}\left[f(k_1)-f(k_2)\right];\quad\langle Q_{DA}\rangle=\frac{(T_2-T_1)}{2}f(k_2); \\
\end{align}
where $G(k)$ is defined such that $G'=f/k$. The maximum efficiency in
the limit $T_1\gg T_2$ then reads
\begin{equation}
{\mathcal E}_\text{sat}=\frac{G(k_1)-G(k_2)}{f(k_1)+G(k_1)-G(k_2)}.
\end{equation}
This leads to the simple criterion that, in this limit, the cycle
outperforms an equilibrium Stirling engine if and only if
\begin{equation}
  \int_{k_2}^{k_1}\frac{f(k)}{k f(k_1)}\dd k > \int_{k_2}^{k_1}\frac{\dd k}{k}.
\end{equation}
In particular, if $f(k)$ is an increasing function of $k$ in the range
$[k_2;k_1]$, the active engine outperforms the equilibrium one. This
could correspond to a physical situation in which energy injection
happens at a particular, finite length scale. As an example, a
semi-flexible filament immersed in a bath of Active Brownian particles
is excited at a characteristic length scale~\cite{nikola_active_2016}
and could thus be a candidate to realize such an efficient engine.

\section{Discussion: back to experiments}
We have assumed all along that the tagged colloidal particle is
subjected to a noise that inherits its properties from those of the
bath while the rest of its dynamics is unchanged. That the effect of
the nonequilibrium bath can be encoded in a single random signal as an
extra force does not seem to be an outrageous hypothesis, though,
given the size of the bacteria used in
\cite{krishnamurthy_micrometre-sized_2016}, comparable to that of the
colloidal particle, perhaps hydrodynamic effects should be taken into
account, as well as further memory effects (say, in the dissipation
kernel). Within that framework, we have shown that with a definition
of the isothermal process based on a iso-"potential energy", we see no
reason for the equilibrium results to be altered in any way. Coming
back to \cite{krishnamurthy_micrometre-sized_2016}, aside from the
limiting efficiency in the high $T_1\gg T_2$ limit, our theoretical
observation is altogether rather consistent with the experiments. We
have suggested an alernative definition of an isothermal process in
which the active temperature is defined through the diffusion constant
of a particle without any external potential. With this definition, in
stark contrast, the efficiency of a Stirling engine takes a
dramatically different form that involves the persistence time of the
noise produced by the bacteria. Interestingly, we are able to pinpoint
memory effects as being responsible for nontrivial efficiencies. Non
Gaussian statistics alone is not a sufficient ingredient (we have
shown equipartition to hold in the limiting non Gaussian but white
scenario).

We hope the suggestion to use our alternative active temperature will
trigger further experiments along the lines of
\cite{krishnamurthy_micrometre-sized_2016}.

\vspace{6pt}

\acknowledgments{The authors thank Julien Tailleur and Jordan Horowitz
  for many discussions. A.S and T.G. acknowledge funding from the
  Betty and Gordon Moore Foundation.}

\appendixsections{multiple} 
\appendix
%

\section{Active particle dynamics}
\label{sec:appActivedyn}
For completeness, we define here Active Brownian and Run-and-Tumble
particles (hereafter ABPs and RTPs). In both cases, the noise entering
the Langevin equation Eq.~\eqref{eq-Langevin} is written as a
force of constant magnitude $f_0$ in a fluctuating direction $\vec u$,
a unit vector. In arbitrary dimension (ABPs are only defined in
$d\ge 2$)
\begin{equation}
  \gamma \dot{\vec r}=-\grad V+\gamma f_0\vec u.
\end{equation}
For ABPs, the direction $\vec u$ undergoes rotational diffusion while
for RTPs a new direction is picked uniformly at a constant rate
$\alpha$. Let us show that, in both cases, each component of the noise has
correlations given by Eq.~(\ref{noiseP}). We focus here on the 2d
case. The derivation follows in the same way in higher dimensions (and
$d=1$ for RTPs).\\ 

In 2d, $\vec u$ is parametrized by an angle $\theta$,
$\vec u=(\cos\theta,\sin\theta)$. For ABPs, the Fokker-Planck equation
associated with the evolution of the angle reads
\begin{equation}
  \label{eq:FP-ABPs}
  \partial_t \cP_t(\theta)=D_r\frac{\partial^2 \cP(\theta)}{\partial\theta^2}
\end{equation}
with $D_r$ the rotational diffusion coefficient. This gives for the
$x$-component of $u_x=\cos\theta$
\begin{equation}
  \partial_t \langle \cos\theta\cos\theta_0\rangle =D_r\int \dd\theta  \cos\theta\cos\theta_0 \frac{\partial^2 \cP(\theta)}{\partial\theta^2}=-D_r\langle\cos\theta\cos\theta_0\rangle
\end{equation}
where the last equality follows from integrating by parts. We thus have
$\langle \cos\theta(t+t_0)\cos\theta(t_0)\rangle=\frac{1}{2}e^{-D_r
  t}$
so that, in the notations of Eq.~(\ref{noiseP}), $\tau=D_r^{-1}$ and
$T=f_0^2/(2D_r)$.

One gets a similar result for RTPs which obey the Master equation
\begin{equation}
  \label{eq:FP-ABPs}
  \partial_t \cP_t(\theta)=-\alpha\cP(\theta)+\alpha\int \frac{\dd\theta'}{2\pi} \cP(\theta')
\end{equation}
with $\alpha$ the tumble rate. This leads in the same way to
$\tau=\alpha^{-1}$ and $T=f_0^2/(2\alpha)$.

\section{Is kurtosis related to efficiency?}
\label{appendix:kurtosis}
One may want to quantify deviations to the Gaussian distribution for the position~\cite{krishnamurthy_micrometre-sized_2016}. We define $\mu_n(X)$ the $n$-th moment of a random variable $X$ and we can compute renormalized kurtosis $\kappa$ defined as follow:
$\kappa_x=\frac{\mu_4(x)}{3\mu_2(x)^2}-1=\frac{\langle x^4\rangle}{3\langle x^2\rangle^2}-1$ if $\langle x\rangle=0$. 
Let us focus on the ABPs case assuming that the derivation of the fourth moment for RTPs is similar. The Fokker-Planck equation in the 2d case for ABPs writes:
\begin{equation}
	\label{eq:fullFP-ABPs}
	\partial_t\cP(\vec r,\theta)=\frac{1}{\gamma}\nabla\cdot(\cP(\vec r,\theta)\nabla V)-f_0\vec u(\theta)\cdot \nabla \cP(\vec r,\theta)+D_r \frac{\partial^2}{\partial \theta^2}\cP(\vec r, \theta).
\end{equation}
We take a quadratic potential $V=\frac 1 2 k \vec r^2$. In the stationary regime, multiplying the two members by $x^4$ and performing integration with respect to $\vec r$ and $\theta$ gives:
\begin{equation}
\label{eq:x4}
	\langle x^4\rangle=f_0 \frac{\gamma}{k} \langle x^3 \cos \theta\rangle=\frac{f_0}{\Omega}\langle x^3 \cos \theta\rangle.
\end{equation}
Similarly we obtain:
\begin{align}
\langle x^3\cos\theta\rangle=\frac{3f_0}{D_r+3\Omega}\langle x^2 \cos^2\theta\rangle;\quad \langle x^2 \cos^2\theta\rangle=\frac{1}{2D_r+\Omega}\left(D_r\langle x^2\rangle+f_0\langle x \cos^3\theta\rangle\right);\\
\langle x \cos^3\theta\rangle=\frac{1}{9D_r +\Omega}\left(\frac{3}{8}f_0+6D_r\langle x\cos\theta\rangle\right);\quad\langle x^2\rangle=\frac{f_0}{\Omega}\langle x\cos\theta\rangle;\quad
\langle x\cos\theta\rangle=\frac{f_0}{2(D_r+\Omega)}.
\label{eq:xcos}
\end{align}
Using Eq.~\eqref{eq:x4} to Eq.~\eqref{eq:xcos}, we get:
\begin{equation}
\kappa_x=-\frac{\Omega(7D_r+3\Omega)}{2(2D_r+\Omega)(D_r+3\Omega)}<0.
\end{equation}
We might wonder whether the kurtosis can give indications on the
efficiency of the stochastic Stirling engine. We can also compute
kurtosis of $x$ for RTPs in 1d as we know the distribution
\cite{0295-5075-86-6-60002}. For this case we have
$\kappa_x=-\frac{2\Omega}{\alpha+3\Omega}$ with $\alpha$ the tumbling
rate. Here $\kappa_x<0$ and we have proved in Sec.~\ref{bath-pers}
that efficiency was still lower than efficiency of the equilibrium
case. This result should be compared to the kurtosis for $x$ that
satisfies the steady state distribution of Sec.~\ref{Sec:nonGaussian}
where the noise is white and non Gaussian. For
$P_{\text{ss}}(x)=C|x/a|^{s}K_s(|x|/a)$ and $C=2^{-s}
a^{-1}/(\sqrt{\pi}\Gamma(1/2+s))$, kurtosis $\kappa_x=2/(1+2s)$ is
strictly positive and the maximum efficiency is still the equilibrium
one. Hence the kurtosis of the position distribution does not indicate
how the efficiency relates to that of an equilibrium Stirling engine.

\bibliographystyle{mdpi} \bibliography{engine-bib}
\end{document}